\documentclass[superscriptaddress, twocolumn, longbibliography, reprint, aps, prl, floatfix]{revtex4-2}
\usepackage{graphicx}              
\usepackage{hyperref}
\usepackage{amsmath}
\usepackage{natbib}
\usepackage{bm}
\usepackage{float}
\usepackage{color}
\usepackage{braket}
\usepackage{ulem}
\usepackage{siunitx}
\usepackage{orcidlink}

\begin{document}
\title{Spin interferometry in a beam of ultracold molecules} 
\author{R. A. Jenkins\orcidlink{0000-0002-6181-9824}}\thanks{These authors contributed equally to this work.}
\author{M. T. Ziemba\orcidlink{0009-0006-0691-1464}}\thanks{These authors contributed equally to this work.}
\author{F. J. Collings\orcidlink{0000-0001-8105-5190}}
\author{X. S. Zheng\orcidlink{0009-0008-4196-5155}}
\author{F. Castellini\orcidlink{0009-0008-1470-1138}}
\author{E. Wursten\orcidlink{0000-0002-2413-2214}}
\author{J. Lim\orcidlink{0000-0002-1803-4642}}
\email{j.lim@imperial.ac.uk}
\author{B. E. Sauer\orcidlink{0000-0002-3286-4853}}
\author{M. R. Tarbutt\orcidlink{0000-0003-2713-9531}}
\email{m.tarbutt@imperial.ac.uk}

\affiliation{Centre for Cold Matter, Blackett Laboratory, Imperial College London, London SW7 2AZ, United Kingdom}

\begin{abstract}
We describe a spin interferometer using ultracold YbF molecules and develop the complete set of techniques needed to measure the electron's electric dipole moment, $d_e$, with this apparatus. The molecules are cooled in an optical molasses and prepared in a single internal quantum state. A Raman transition prepares a spin superposition which evolves in parallel magnetic and electric fields before a second Raman transition maps the phase onto the populations of two hyperfine states. These populations are read out using detectors that have spatial and temporal resolution and approach unit efficiency. We characterize the efficiencies and fidelities of all these steps and evaluate the sensitivity of this approach to measuring $d_e$.
\end{abstract}

\maketitle 

{\it Introduction}---In the Standard Model (SM) of particle physics, the predicted value of the electron's electric dipole moment (eEDM) is $d_e \approx 1 \times  10^{-35}$ $e\, \text{cm}$~\cite{Ema2022}. Many extensions of the SM, including supersymmetric extensions, introduce new CP-violating interactions producing eEDM values orders of magnitude larger~\cite{Engel2013, Nakai2017, Chupp2019, Safronova2018}. These theories address inadequacies of the SM and may help explain the matter-antimatter asymmetry of the Universe~\cite{Dine2003, Sakharov1991}, so it is important to seek experimental evidence that supports them. So far, no eEDM has been detected, and the current upper limit is $d_e < 4.1 \times 10^{-30}$ $e\, \text{cm}$ (90\% confidence)~\cite{Roussy2023}. This is a sensitive probe of physics beyond the SM that constrains broad classes of new physics at energies exceeding  10~TeV.

The eEDM is determined by measuring the electron spin precession frequency in an electric field. The most precise measurements use electrons bound inside heavy polar molecules, exploiting the enormous effective electric fields, $E_{\rm eff} \sim 10-100$~GV/cm, offered by these species. The statistical sensitivity of these measurements is proportional to $E_{\rm eff} \tau \sqrt{n}$, where $\tau$ is the spin precession time and $n$ is the number of molecules detected. Some experiments use molecular beams of neutral molecules~\cite{Hudson2011, Andreev2018}, where $n$ can be large but $\tau$ is limited to about a millisecond by the beam divergence. Another approach uses trapped molecular ions~\cite{Zhou2020, Roussy2023}, where $\tau$ can be many seconds, but $n$ is limited by the Coulomb interactions. A newer idea, stimulated by recent advances in laser cooling and trapping of molecules~\cite{Fitch2021b, Augenbraun2023}, is to reach large $n$ and $\tau$ by using ultracold neutral molecules in a beam or optical trap~\cite{Fitch2020b, Aggarwal2018, Augenbraun2020, Zeng2024, Bause2025, Takahashi2025arxiv}. The principles of an eEDM measurement with optically trapped molecules have been demonstrated with CaOH~\cite{Anderegg2023}, but it has only a small $E_{\rm eff}$. A few species with large $E_{\rm eff}$ have been laser-cooled, most notably YbF~\cite{Lim2018, Alauze2021}, BaF~\cite{Rockenhauser2024, Zhang2022c}, YbOH~\cite{Augenbraun2020} and SrOH~\cite{Lasner2025}, but the utility of these ultracold molecules for eEDM measurements has not previously been shown. Here, we develop the complete set of techniques needed to measure $d_e$ using a beam of laser-cooled YbF molecules.

\begin{figure*}[tb]
    \centering
    \includegraphics[width=\linewidth]{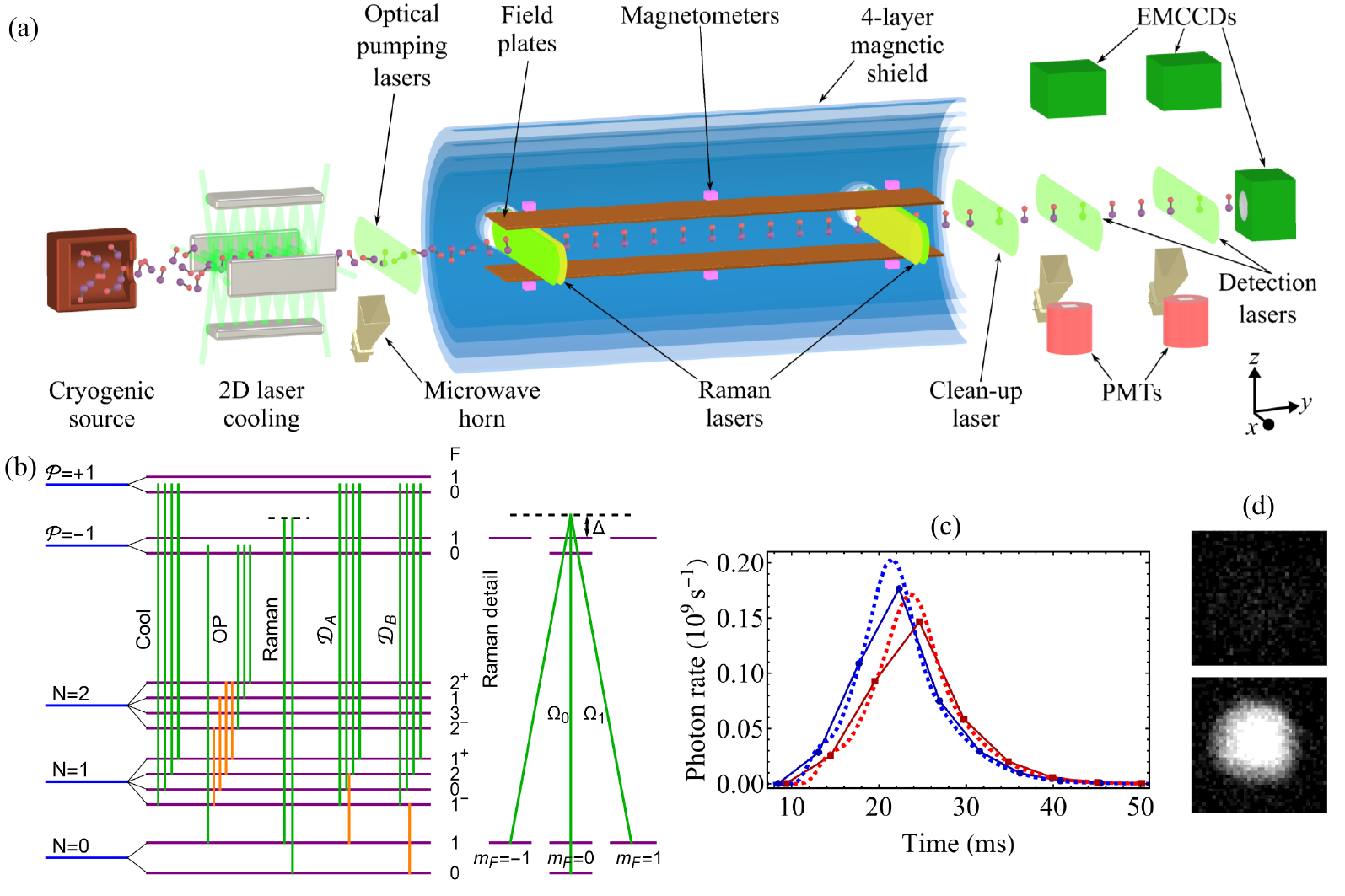}
    \caption{(a) Illustration of the experiment. (b) Relevant energy levels of $^{174}$YbF, together with lasers (green) and microwaves (orange) used in the experimental sequence. The lower levels are rotational ($N$) and hyperfine ($F$) states within $X^{2}\Sigma^{+} (v=0)$. The upper levels are the two parity (${\cal P}$) components of $A^{2}\Pi_{1/2}(v=0,J= 1/2)$. From left to right the steps are: laser cooling; optical pumping; Raman transfer; detector A; detector B; detail showing the $(F,m_F)$ levels coupled by Raman lasers. (c) Time-of flight profiles recorded by EMCCDs (points with solid lines) and PMTs (dashed lines) in the first (blue) and second (red) detectors. Vertical axis applies to EMCCD data. PMT data is scaled by the ratio of detection efficiencies to make it visible on the same scale. (d) Images from the end-on camera for laser cooling off (upper) and on (lower). }
    \label{fig:experimentsetup}
\end{figure*}

{\it Overview}---Figure \ref{fig:experimentsetup}(b) shows the relevant energy levels of $^{174}$YbF and the transitions driven at each step of the experiment. We use $\ket{N,F,m_F}$ to denote levels within the ground electronic and vibrational state $X^{2}\Sigma^{+} (v=0)$, and the shorthand notation $\ket{0}=\ket{0,0,0}$, $\ket{x} =\frac{1}{\sqrt{2}}\left(\ket{0,1,+1} + \ket{0,1,-1} \right)$  and $\ket{y}=\frac{i}{\sqrt{2}}\left(\ket{0,1,+1} - \ket{0,1,-1} \right)$. We use $\ket{\{N\}}$ for the set of components of rotational state $N$ and $\ket{\{N,F\}}$ for all components of a hyperfine state. The electronic excited state $A^{2}\Pi_{1/2}(v=0,J= 1/2)$ is denoted $\ket{e\pm}$, where $\pm$ refers to the parity. 

Figure \ref{fig:experimentsetup}(a) illustrates the experiment, which can be understood as a molecular spin interferometer. Every 200~ms, a pulse of YbF with a mean speed of 170~m/s is emitted from a 3.7~K cryogenic buffer gas source~\cite{Truppe2017c}. The molecules are cooled in an optical molasses to $\sim100~\mu$K in both transverse directions~\cite{Alauze2021}, producing a highly collimated beam. They are optically pumped to $\ket{0}$, then enter a magnetically-shielded region~\cite{Collings2025} where they are subjected to uniform electric and magnetic fields $(E,B)\hat{z}$. Once in $E$, they are transferred to $\ket{y}$ using a stimulated Raman process which forms the splitter of the interferometer. This state evolves for a time $\tau$ to $\sin\phi\ket{x} + \cos\phi\ket{y}$, where $\phi =\frac{1}{\hbar} \int_{0}^{\tau} \Delta E(t)\,dt$. Here, $\Delta E$ is half the energy difference between $m_F=\pm 1$, $\Delta E = \mu_{\rm B}B -d_e E_{\rm eff}$ where $E_{\text{eff}} = E_{\text{eff}}^{\text{max}} \eta(E)$, $E_{\text{eff}}^{\text{max}} = -26$~GV/cm and $\eta(E)$ is the polarization factor which is 0.693 at our operating field of $E=20$~kV/cm. After travelling $L=0.77$~m, the molecules encounter a second pair of Raman beams which form the recombiner, mapping $\ket{y} \leftrightarrow \ket{0}$ but doing nothing to $\ket{x}$. Finally, the populations $P_{1,0}$ in $F=1,0$ are measured using two consecutive detectors, ${\cal D}_{A,B}$ whose efficiencies are $\epsilon_{A,B}$. In the detectors, laser induced fluorescence (LIF) is imaged onto photomultiplier tubes (PMTs) and cameras (EMCCDs), yielding the signals $S_{A,B}$. Figure \ref{fig:experimentsetup}(c) shows example time-of-flight profiles recorded by the detectors.  A third camera at the end of the beamline measures the density distribution in the $xz$-plane, as illustrated in Fig.~\ref{fig:experimentsetup}(d).  We form the asymmetry ${\cal A}=(S_B - \frac{\epsilon_B}{\epsilon_A} S_A)/(S_B+ \frac{\epsilon_B}{\epsilon_A}S_A)$ where $\epsilon_B/\epsilon_A$ is included to remove the dependence on the relative detector efficiencies which we measure separately. In the ideal experiment, $P_1=\sin^2\phi$, $P_0=\cos^2\phi$, $\epsilon_{B}/\epsilon_{A}=1$ and ${\cal A}=\cos(2\phi)$. More details about the experimental setup are given in Appendix A. Here, we characterize each of the control and readout steps.

{\it Laser cooling}---In the 2D optical molasses, which is 0.2~m long and centred 0.70~m from the source, we use the magnetically-assisted Sisyphus effect to cool the molecular beam to sub-Doppler temperatures, following the methods described in \cite{Alauze2021}. The main laser cooling light, labelled `Cool' in Fig.~\ref{fig:experimentsetup}(b), is blue-detuned from the transition $\ket{e+}\leftrightarrow \ket{\{1\}}$, repump lasers are used to recover population that leaks to the higher-lying vibrational levels, $v=1,2$, and radiofrequency (rf) sidebands are applied to all cooling lasers to address the hyperfine structure. The effectiveness and accurate alignment of the laser cooling is monitored by imaging LIF onto a camera 4.88~m from the source. Figure \ref{fig:experimentsetup}(d) shows  images from this end-on camera when the laser cooling is off (top) and on (bottom). The cooling is most effective for slower molecules, and increases the flux of molecules below 150~m/s by a factor of about 10. 

\begin{figure}[tb]
    \centering
    \includegraphics[width=\linewidth]{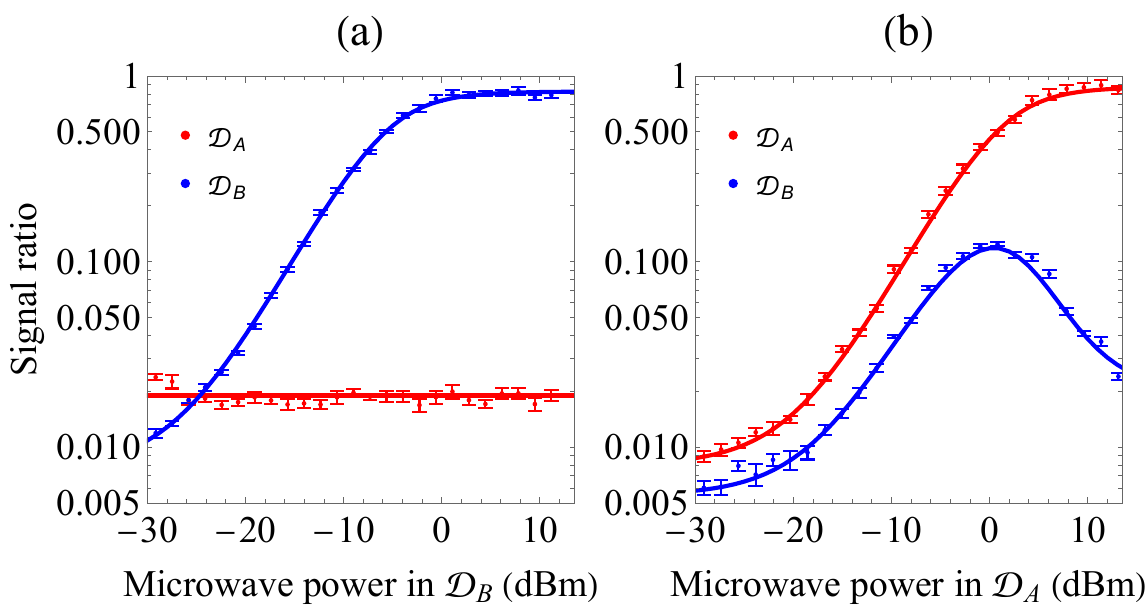}
    \caption{Detector characterization, showing efficiency of measuring $P_{0,1}$ as a function of microwave powers, $\Pi^{\rm MW}_{A,B}$. Signals are normalized to those obtained without optical pumping [population remains in $N=1$]. (a) Signal ratios as a function of $\Pi^{\rm MW}_{B}$, with molecules prepared in $F=0$ and $\Pi^{\rm MW}_{A}=12$~dBm. (b) Signal ratios as a function of $\Pi^{\rm MW}_{A}$ with molecules prepared in $F=1$ and $\Pi^{\rm MW}_{B}=5$~dBm. Points: $r_A$ (red) and $r_B$ (blue). Lines: model described in Appendix B.}
    \label{fig:detectors}
\end{figure}

{\it Optical pumping}---Laser cooling proceeds from $N=1$, but the eEDM experiment uses $N=0$. Thus, after cooling, we optically pump the molecules from all 12 components of $\ket{\{1\}}$ into $\ket{0}$, using the scheme labelled `OP' in Fig.~\ref{fig:experimentsetup}(b). A 29~GHz microwave field, with rf sidebands, connects $\ket{\{1\}} \leftrightarrow \ket{\{2\}}$, and lasers connect $\ket{\{2\}} \leftrightarrow \ket{e-}$ and $\ket{\{0,1\}} \leftrightarrow \ket{e-}$. This excited state can decay back to the bright states (those addressed by the lasers) with probability $b_{\rm b}=\frac{7}{9} b_{00}$, or to the target state $\ket{0}$ with probability $b_{\rm t}=\frac{2}{9}b_{00}$, or to higher-lying vibrational states with probability $1-b_{00}$, where $b_{00} = 0.933(3)$ is the relevant vibrational branching ratio~\cite{Zhuang2011}. Thus, ideally, the fraction of the cooled population transferred to $\ket{0}$ is $\epsilon_{\rm OP}=b_{\rm t}/(1-b_{\rm b})=0.756(9)$. All the rest should be pumped to $v=1$. The OP is not perfect, and some molecules in $v=1$ decay back to $v=0$ between the OP and detection regions, producing a background $P_{\rm bg}$. To mitigate this, a `clean up' laser beam drives transitions from $\ket{\{1\}}$ to $A^{2}\Pi_{1/2}(v=0,J=3/2)$. We measure both $\epsilon_{\rm OP}$ and $P_{\rm bg}$ below. 

\begin{figure}[tb]
    \centering
    \includegraphics[width=\linewidth]{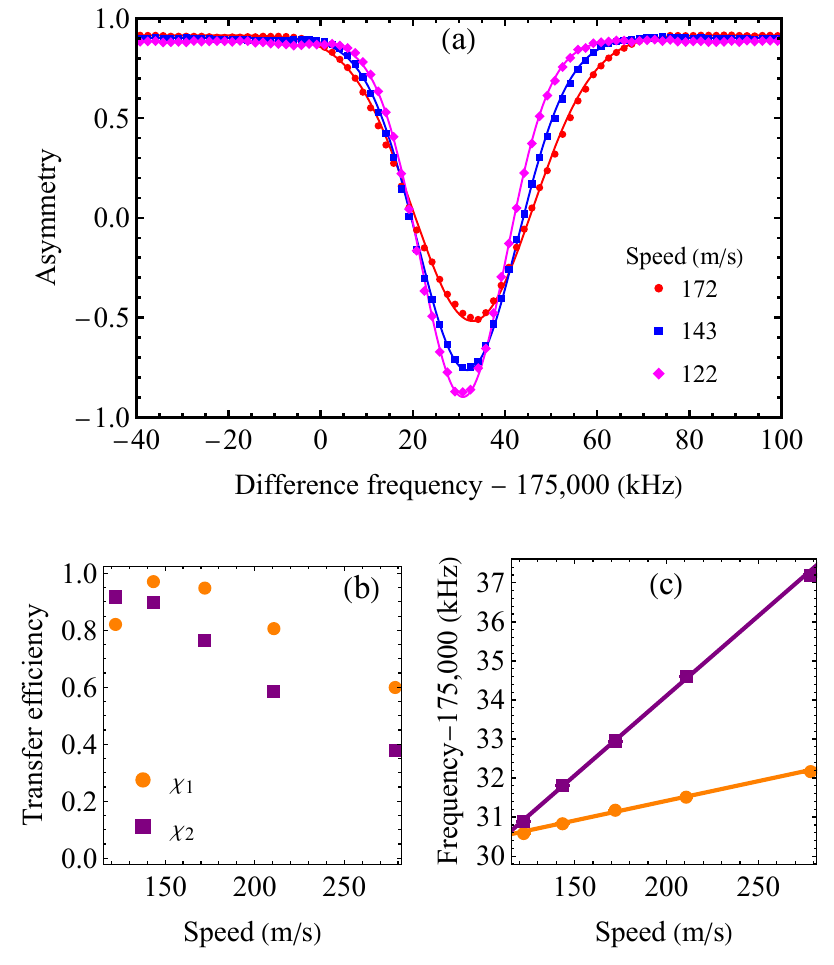}
    \caption{(a) Raman transfer in the recombiner, for three different velocities, plotted as asymmetry versus two-photon detuning. Data is from EMCCDs.  Parameters are $\Delta = -9.55\times 10^9$~rad/s and $\Omega_{0,1}=5.7 \times 10^7$~rad/s. Points: data. Lines: Gaussian fits. (b) Transfer efficiency in the splitter ($\chi_1$, orange circles) and recombiner ($\chi_2$, purple squares). (c) Doppler shifts in the two regions, with linear fits. }
    \label{fig:stirap}
\end{figure}

{\it Detection}---We measure the populations $P_{1,0}$ with high efficiency, low background and minimal cross-talk using the detection schemes labelled ${\cal D}_{A,B}$ in Fig.~\ref{fig:experimentsetup}(b). We selectively couple the $N=0$ hyperfine components to $N=1$ using resonant microwaves, drive the optical cycling transition $\ket{\{1\}} \leftrightarrow \ket{e+}$, and detect the resulting fluorescence with PMTs and EMCCDs. The efficiency of detecting a molecule this way is approximately 4.5\% for the PMTs and 54\% for the EMCCDs. The EMCCDs measure about $10^6$ photons per shot. Figure \ref{fig:detectors} studies the performance of the detectors as a function of the microwave powers at the detector horns, $\Pi^{\rm MW}_{A,B}$.  To determine the effectiveness of the detectors and the optical pumping, we interleave measurements of $S_{A,B}$ with and without OP, then form the ratio $r_{i}=S_{i}^{\rm on}/S_{A}^{\rm off}$ where on and off refer to the OP microwaves. In Fig.~\ref{fig:detectors}(a), we prepare molecules in $F=0$ and measure $r_{A,B}$ as a function of $\Pi^{\rm MW}_{B}$. Ideally, $F=0$ population is measured exclusively in ${\cal D}_B$. We see that $r_{B}$ increases with power up to about 2~dBm where it saturates at 80(1)\%. After correcting for small background $N=0$ and $N=2$ populations which we measure separately, we determine $\epsilon_{\rm OP}=0.738(11)$, consistent with the ideal efficiency given above. There is a 1.9\% background in $r_A$ due to residual population in $N=1$ (0.8\%) and $F=1$ (1.1\%). Defining $P_{\rm bg}$ to be the background population divided by $P_0$, we find $P_{\rm bg}=2.33(4)$\%. The background does not change with $\Pi^{\rm MW}_{B}$ showing that microwave leakage between the detectors is small. We infer an upper limit to microwave-leakage-induced cross talk of ${\rm x}_A<0.002$. In Fig.~\ref{fig:detectors}(b), optical pumping prepares molecules in $F=1$ and we measure $r_{A,B}$ as a function of $\Pi^{\rm MW}_{A}$. As the power increases, $r_A$ rises from a background of 0.8\% (residual $N=1$ population), and saturates to 83(2)\% at powers above 8~dBm.  At very low powers, the $F=1$ molecules pass through ${\cal D}_A$ without scattering, so remain in $F=1$ and are not visible in ${\cal D}_B$. At intermediate powers, molecules leave ${\cal D}_A$ before they are pumped to $v=1$, and the fraction of this population transferred to $N=1$ increases with $\Pi^{\rm MW}_{A}$. That fraction is detected in ${\cal D}_B$, resulting in cross-talk between the two detectors which initially increases with $\Pi^{\rm MW}_{A}$ (blue points in Fig.~\ref{fig:detectors}(b)). At high power, most molecules are pumped to $v=1$ in ${\cal D}_A$ so $r_B$ decreases again. At our operating power of 10~dBm, the cross-talk due to incomplete detection is ${\rm x}_B=0.030$. We model these results using the four-level rate model described in Appendix B and illustrated in Fig.~\ref{fig:detector_model}. The results of this model fit very well to the data, as can be seen by the lines in Fig.~\ref{fig:detectors}. To get a good fit, we have set the excitation rate at the centre of the laser beam to $R_{1{\rm e}}^{\rm max}=2.5 \times 10^6$\,s$^{-1}$, and the microwave coupling rates to $R_{01}=0.1 \times 10^6$\,s$^{-1}$ when $\Pi^{\rm MW}_{A}=10$~dBm and $R_{01}=0.49 \times 10^6$\,s$^{-1}$ when $\Pi^{\rm MW}_{B}=10$~dBm.

{\it Raman transfer}---The Raman scheme is shown in Fig.~\ref{fig:experimentsetup}(b). A pair of laser beams, co-propagating along $x$, with Rabi frequencies $\Omega_{0,1}$, difference frequency $\delta$ and single-photon detuning $\Delta$, drive the $\ket{0}\leftrightarrow \ket{y}$ transition. The beam addressing $F=0$ ($F=1$) is linearly polarized along $z$ ($y$), so that $\ket{0}$ is coupled to $\ket{y}$ via the $(F,m_F)=(1,0)$ component of $\ket{e-}$, whereas $\ket{x}$ is not coupled. The beams can be separated along $y$ to explore Stimulated Raman Adiabatic Passage, but in the present work they are (nearly) overlapped so that we drive a $\pi$-pulse for molecules near the mean velocity. Figure \ref{fig:stirap}(a) shows the transfer efficiency of the recombiner alone (the splitter is turned off). Very similar results are obtained for the splitter. The molecules are prepared in $\ket{0}$ and we measure the asymmetry as a function of $\delta$ using time-resolved data from the EMCCDs to select three different velocities. We fit Gaussians to these data to determine the transfer efficiencies and centre frequencies. Figure \ref{fig:stirap}(b) shows the splitter and recombiner efficiencies, $\chi_1$ and $\chi_2$, for various speeds. The efficiency is high for molecules around the mean speed, but drops off for higher speeds. The weighted mean values~\footnote{Each velocity bin is assigned a weight proportional to its eEDM sensitivity, i.e. to $\tau\sqrt{n}$} are $\langle\chi_1\rangle=0.88$, $\langle\chi_2\rangle=0.76$. Further analysis of the Raman efficiency is given in Appendix C. Figure \ref{fig:stirap}(c) shows the centre frequencies for different speeds. The variation is the Doppler shift arising from a small angle between the two Raman beams in each region. These shifts are $10.1(5)$~Hz/(m/s) in the splitter, and $40.9(6)$~Hz/(m/s) in the recombiner, corresponding to angles of $\alpha_1=5.6(3)$~$\mu$rad and $\alpha_2=22.6(3)$~$\mu$rad. Extrapolating to zero velocity, the centre frequencies between splitter and recombiner differ by $3.48(14)$~kHz due to the difference in their Stark shifts. At $E=20$~kV/cm, the sensitivity is 0.152~kHz/(V/cm), so the observed difference corresponds to a change in plate spacing of $20.8(8)$~$\mu$m between the two regions. As in previous eEDM measurements~\cite{Hudson2011,Kara2012}, we will introduce small steps in the Raman frequencies so that $\delta$ remains centred at zero even if there are drifts in the Doppler shifts.

\begin{figure}[tb]
\centering
\includegraphics[width = \linewidth]{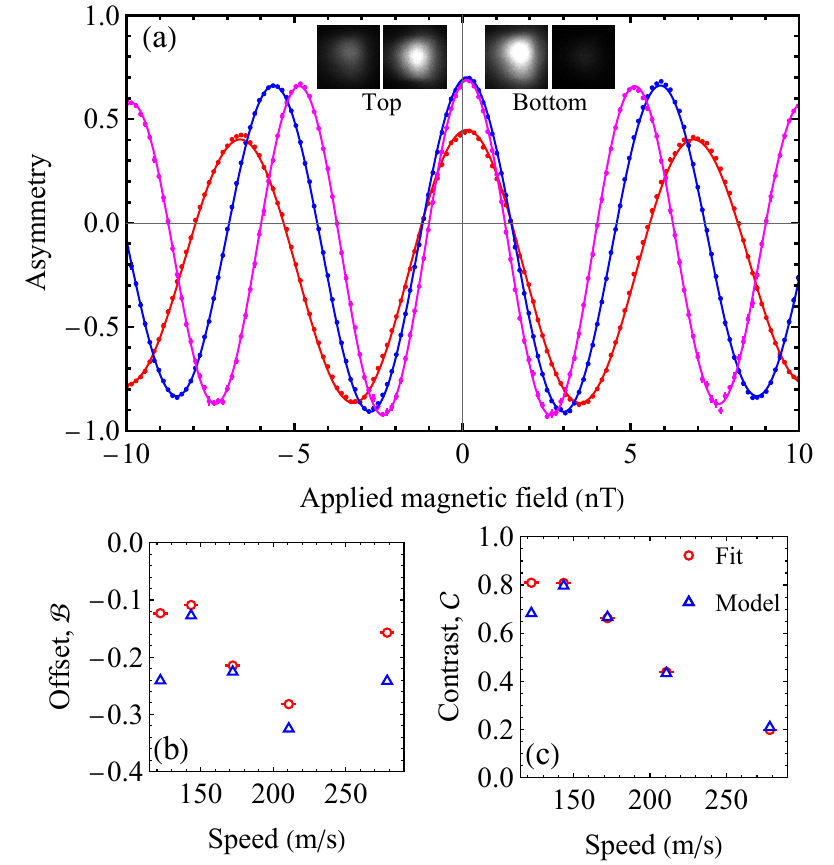}
\caption{(a) Interference fringes for the same three velocities as Fig.~\ref{fig:stirap}. Points: data. Lines: Fit to Eq.~(\ref{eq:interference_model_averaged_evaluated}). Inset: EMCCD images in ${\cal D}_A$ and ${\cal D}_B$ at the top and bottom of the fringe. (b, c) Fit parameters ${\cal B, C }$ for various speeds (red circles) compared to the values expected from the model in Appendix C (blue triangles).}
\label{fig:interference}
\end{figure}

{\it Interferometer}---Figure \ref{fig:interference}(a) shows typical interferometry results obtained from EMCCD data. We plot the asymmetry ${\cal A}$ as a function of the applied magnetic field, $B$, for the same three velocities as in Fig.~\ref{fig:stirap}. We fit the data to Eq.~(\ref{eq:interference_model_averaged_evaluated}) with the offset ${\cal B}$, contrast ${\cal C}$ and background magnetic field $B_{\rm bg}$ as free parameters. Their weighted mean values~\cite{Note1} are $\langle{\cal B}\rangle=-0.18$, $\langle{\cal C}\rangle=0.65$ and $\langle B_{\rm bg} \rangle = 123$~pT. The red circles in Fig.~\ref{fig:interference}(b,c) show the fitted values of ${\cal B}$ and ${\cal C}$ for various speeds. The contrast reaches 0.8 at low speeds, but drops off for higher speeds. The blue triangles in the figure show the values of ${\cal B}$ and ${\cal C}$ predicted by the model presented in Appendix C and our measured values of Raman efficiencies ($\chi_{1,2}$), detector cross talk (${\rm x}_{A,B}$) and background $P_{\rm bg}$. The agreement is good for most speeds, but less good for low and high speeds where the number of molecules is small. We record interference curves using PMTs simultaneous with EMCCDs. An analysis of the corresponding PMT data is presented in Appendix C. 

{\it Discussion and outlook}---At the shot noise limit, the statistical uncertainty of the measurement is $\sigma_{d_e}=\hbar/(2{\cal C} E_{\rm eff} \tau \sqrt{n})$. With the values demonstrated here (${\cal C}=0.65$, $\tau=5$~ms, $n=2.0\times 10^6$ molecules per shot, 5 shots per second), and assuming a 50\% duty cycle to account for the settle time following each $E$-reversal and the time for auxiliary measurements needed to determine systematic shifts and uncertainties, this limit is $\sigma_{d_e} = 8.6\times 10^{-30}$~e~cm in 24 hours. Thus, using the current apparatus, a measurement with a statistical uncertainty below $10^{-30}$~e~cm is feasible with about 100 days of data. While it is challenging to reach the shot noise limit at this level, we have already demonstrated that magnetic noise is several times smaller than our projected shot noise limit~\cite{Collings2025}, and our detection method is relatively immune to shot-to-shot fluctuations in molecule flux. 
Systematic errors are also very challenging at this level. A primary systematic error is a magnetic field that correlates with the $E$-reversal. We are able to measure this effect with a precision of $3.6 \times 10^{-31}$~e~cm in about 100 hours~\cite{Collings2025}.

The temperature of the laser-cooled beam is 50-500~$\mu$K, depending on the parameters of the molasses~\cite{Alauze2021}. At these temperatures, $\tau$ can be at least 10 times longer than in the present work without significant expansion of the beam size. To increase $\tau$, we have developed a new cryogenic source with a similar flux to the present source, but with a velocity three times smaller~\cite{White2024}, and we have slowed these beams to very low velocity using radiation pressure~\cite{Athanasakis-Kaklamanakis2025}. We have shown that lowering the velocity increases the effectiveness of the molasses, improves the detection efficiency and reduces detector cross talk. The design of the spin precession region is modular~\cite{Collings2025} so that it can be extended, further increasing $\tau$. Even longer $\tau$ may be achieved by trapping the ultracold molecules in an optical lattice~\cite{Fitch2020b} and all the techniques developed here can be applied to such an experiment. Increases in $n$ can come by pumping population from other rotational states into $N=1$ prior to laser cooling, and by focussing molecules from the cryogenic source into the molasses~\cite{Fitch2020b}. With these methods, uncertainties below $10^{-31}~e$~cm could be reached. The methods developed in this work can be applied directly to other eEDM measurements using ultracold molecules in a beam or trap~\cite{Aggarwal2018, Augenbraun2020, Anderegg2023, Zeng2024, Bause2025, Takahashi2025arxiv}. By extending the methods to $^{171}$YbF and $^{173}$YbF, our methods could also be used to probe P,T-violating effects in the nucleus via the nuclear Schiff moment and the nuclear magnetic quadrupole moment~\cite{Grasdijk2021,Ho2023,DeMille2024}. 

    We are grateful to Jon Dyne and David Pitman for their expert technical support and to Noah Fitch, Xavier Alauze, Simon Swarbrick and Gen Li for their earlier contributions. This work has received funding from UKRI under grants EP/X030180/1, ST/V00428X/1, ST/Y509978/1 and MR/Z505122/1, and the European Union's Horizon 2020 research and innovation programme under the Marie Skłodowska-Curie grant agreement No 101212437. The research has been supported by the ``Table-top experiments for fundamental physics'' program, sponsored by the Gordon and Betty Moore Foundation (GBMF12327), Simons Foundation, Alfred P. Sloan Foundation (G-2023-21035), and John Templeton Foundation.

The data that support the findings of this article are openly available~\cite{zenodoYbFInterferometer2026}.

\bibliography{references}

\onecolumngrid
\appendix

\section{End Matter\label{appendix}}

\twocolumngrid

\appendix
\refstepcounter{section}
\setcounter{equation}{0}
\renewcommand{\theequation}{\Alph{section}\arabic{equation}}

{\it Appendix A: Experimental details}---The experiment begins with a cryogenic buffer gas source similar to the one described in \cite{Truppe2017c}. Here, YbF molecules are produced by laser ablation of a Yb target in the presence of SF$_6$ inside a copper cell cooled to 3.7~K. They are entrained in a stream of cold helium gas which flows through the cell at a rate of 1~sccm. The ablation energy is 40~mJ and the repetition rate is 5~Hz. This pulsed molecular beam is then cooled using the magnetically-assisted Sisyphus effect in a two-dimensional optical molasses, following the methods described in \cite{Alauze2021}. The main cooling light drives the optical cycling transition at 552~nm, illustrated in Fig.~\ref{fig:experimentsetup}(b). We set the laser frequency detuned by 34~MHz from the $F=1^-$ component of the cooling transition, then divide the light into two equal parts, shift one part by -159~MHz, then recombine them. The laser has a power of 500~mW. Population leaking to $v=1$ and $v=2$ is recovered by repump lasers at 568~nm and 565~nm, each with appropriate rf sidebands to address the hyperfine structure. To make the molasses, we combine the lasers into a single beam with a waist of 2.5~mm, divide this beam into two parts, and then bounce each back and forth to form two sheets of light parallel to $x$ and $z$, each 20~cm long in the $y$ direction. All the light is linearly polarized along $y$ and a magnetic field of about $100\mu$T is applied in the $xy$-plane at about 45$^{\circ}$ to $y$. The molasses is followed by a short region containing only repump light to ensure that all population is pumped back to $v=0$.

The molecules are optically pumped to $\ket{0}$ using the scheme labelled OP in Fig.~\ref{fig:experimentsetup}(b). We find that two microwave components near 29~GHz, spaced by 20~MHz, each with power of 24~dBm at the horn, are sufficient to couple all hyperfine components of $N=1$ to selected hyperfine components of $N=2$. A laser with a power of 240~mW drives $\ket{\{2\}} \leftrightarrow \ket{e-}$ and another laser with a power of 2~mW drives $\ket{\{0,1\}} \leftrightarrow \ket{e-}$. The optical pumping laser beam has a waist of 4.9~mm along $z$ and 0.75~mm along $y$ and bounces back and forth 5 times to increase interaction time. The molecules scatter photons until they reach $\ket{0}$ or a higher-lying vibrational level. Imperfect optical pumping and vibrational decay of molecules between pumping and detection produce a background of $P_{\rm bg} \approx 0.06$ in the detectors. We reduce this background using the `clean up' laser that drives transitions from $N=1$ to $A^{2}\Pi_{1/2}(v=0,J=3/2)$. This has the effect of transferring the background population to $N=3$.

In the interaction region, described in detail in \cite{Collings2025}, electric field plates, 1.1~m long, produce $E = 20$~kV/cm along $z$. The plates are housed inside a glass vacuum tube surrounded by atomic magnetometers and enclosed in a four-layer magnetic shield. A uniform magnetic field is applied using current-carrying wires inside the inner shield. The splitter and recombiner of the interferometer are implemented by driving Raman transitions in the electric field. The transitions are driven by a pair of laser beams propagating along $x$ with powers of 45~mW and $1/e^2$ diameters of 14.5~mm along $z$ and 4.5~mm along $y$. The beams are roughly Gaussian along $y$ and are separated by 0.7~mm. One beam is linearly polarized along $z$, and detuned by $\Delta$ from the Stark-shifted frequency of $\ket{0} \leftrightarrow \ket{e-}$. The other is linearly polarized along $y$ and  detuned by $\Delta + \delta$ from the Stark-shifted frequency of $\ket{0,1,\pm 1} \leftrightarrow \ket{e-}$, with $\delta \approx 0$. Their peak Rabi frequencies, $\Omega_{0,1}$, are calibrated by measuring Rabi oscillations as a function of $\Delta$. We had initially intended to drive stimulated Raman adiabatic passage but this is problematic in our system due to the close spacing of the ground-state hyperfine levels. Off-resonant cross-couplings result in undesirable ac Stark shifts and excitation of the molecule, limiting the transfer efficiency, and the spatial variation of these shifts interfere with proper adiabatic following. Instead, we drive a $\pi$-pulse for molecules with velocity close to the mean and avoid spontaneous emission by choosing large $|\Delta|$.

In the two detectors, we use microwaves to selectively couple the $F=0$ and $F=1$ components of $N=0$ to $N=1$, and then detect the $N=1$ population by driving the cycling transition. Specifically, we couple $\ket{\{0,1\}}\leftrightarrow \ket{\{1,2\}}$ in ${\cal D}_{A}$ and $\ket{0}\leftrightarrow \ket{\{1,1^-\}}$ in ${\cal D}_{B}$, while in both detectors a laser with appropriate rf sidebands drives $\ket{e+}\leftrightarrow\ket{\{1\}}$. The microwave power delivered to the horns is 10~dBm in both detectors and the laser has a power of 22~mW and a waist of 5.0~mm.  Dark states are eliminated using counter-propagating laser beams and by modulating the polarization of the light at 1~MHz. In each detector, the resulting fluorescence is imaged onto a PMT with a time resolution of 0.1~ms, and onto an EMCCD with $2^{14}$ pixels and a time resolution of 4.62~ms. A sandwich of perforated aluminium and microwave absorbing foam~\footnote{Laird Eccosorb HR}, with a central 10~mm diameter tube for the molecules to pass through, blocks microwave leakage between the two detectors, and more foam inside the chambers reduces reflections from the chamber walls. These steps reduce microwave-induced cross-talk between the detectors to an acceptable level. As the molecules pass through the detector, they are optically pumped to $v=1$. From the results presented in Fig.~\ref{fig:detectors}, and the rate model presented below, we estimate that they scatter  an average of 13.4 photons. We use ray tracing together with the fluorescence images recorded by the cameras to estimate the photon collection efficiency. Separately, we calibrate the quantum efficiencies and gains of the PMTs and EMCCDs. Using these data, we estimate that molecules are detected with an efficiency of 54\% by the EMCCDs and 4.5\% by the PMTs. We measure the relative efficiency of ${\cal D}_A$ and ${\cal D}_B$, $\epsilon_A/\epsilon_B$, by preparing all molecules in $\ket{0}$ then measuring $P_0$ separately in each detector, switching rapidly between the two detectors to average away the shot-to-shot fluctuations of the source. For the data presented in this paper, we find (weighted means) $\langle\epsilon_B/\epsilon_A\rangle=0.89$ for the EMCCDs, and $\langle\epsilon_B/\epsilon_A\rangle=0.68$ for the PMTs.

\refstepcounter{section}
\setcounter{equation}{0}
\renewcommand{\theequation}{\Alph{section}\arabic{equation}}

\begin{figure}[tb]
\centering
\includegraphics[width = 0.6\linewidth]{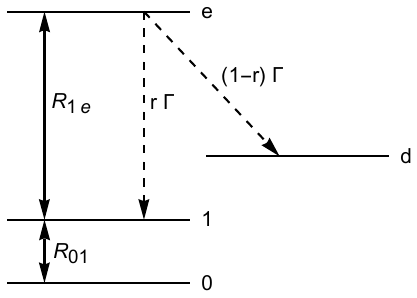}
\caption{Rate model to describe detectors.}
\label{fig:detector_model}
\end{figure}

\par\noindent\phantomsection\label{app:detectorModel}
{\it Appendix B: Modelling detectors}---We describe the detectors using the four-level rate model illustrated in Fig.~\ref{fig:detector_model}. Levels 0 and 1 are coupled by microwaves at rate $R_{01}$, while 1 and e are coupled by a laser at rate $R_{1{\rm e}}$. The excited state e can decay back to 1 or to a dark state d, with branching ratios $r$ and $1-r$ and total decay rate $\Gamma$. The rate equations for the populations $N_{0,1,{\rm e},{\rm d}}$ are:
\begin{align}
\dot{N}_0 &= R_{01}(N_1 - N_0), \\
\dot{N}_1 &= -R_{01}(N_1 - N_0) +R_{1{\rm e}}(N_{\rm e} - N_1) + \Gamma r N_{\rm e},\\
\dot{N}_{\rm e} &= -R_{1{\rm e}}(N_{\rm e} - N_1) - \Gamma N_{\rm e}, \\
\dot{N}_{\rm d} &= \Gamma(1-r)N_{\rm e}.
\end{align}
We solve these rate equations for molecules passing through the detector and calculate the fluorescence signal as $S=\Gamma \int N_{\rm e}(t)\,dt$. Any population remaining in state 1 in the first detector is made available in the second detector. For these simulations, we fix $r=0.93$ and $\Gamma = 35.8\times 10^6$~s$^{-1}$~\cite{Zhuang2011}, take a uniform microwave coupling, assume $R_{01}$ is proportional to microwave power, and set $R_{1e}(t)=R_{1{\rm e}}^{\rm max}e^{-2t^2/\tau^2}$ with $\tau=33$~$\mu$s, corresponding to the transit time through the probe laser for molecules at the mean speed. To obtain the signal ratios $r_{A,B}$ we divide the signal due to an $N=0$ molecule (OP on) by that of an $N=1$ molecule (OP off), then multiply by the ideal optical pumping efficiency along with a small correction due to the presence of $N=0$ and $N=2$ molecules in the beam. Backgrounds due to molecules entering in the wrong states are accounted for in the model. We choose $R_{01}$ and $R_{1e}^{\rm max}$ so that the simulations give a good match to experimental results.

\refstepcounter{section}
\setcounter{equation}{0}
\renewcommand{\theequation}{\Alph{section}\arabic{equation}}

\begin{figure*}[tb]
\centering
\includegraphics[width = 0.85\linewidth]{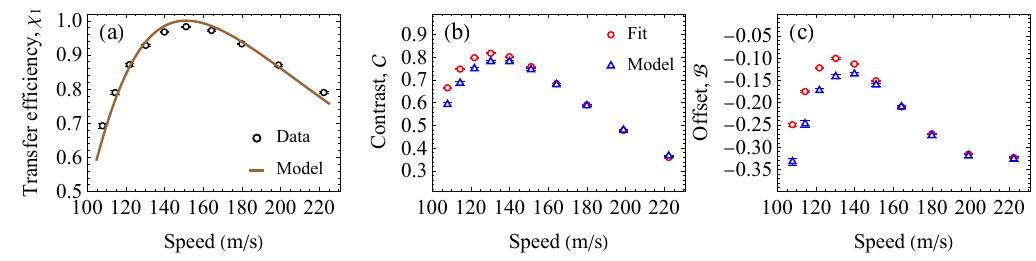}
\caption{(a) Raman efficiency, $\chi_1$, as a function of molecular speed. Points: measurements using PMT data. Line: Analytical model. (b) Contrast, ${\cal C}$, and (c) offset ${\cal B}$ of interference curves measured using PMTs for molecules of various speeds. Red circles are determined by fitting Eq.~(\ref{eq:interference_model}) to the data (we set ${\cal D}=0$). Blue triangles are from Eqs.~(\ref{eq:interference_contrast}) and (\ref{eq:interference_background}).}
\label{fig:pmt_fits}
\end{figure*}

{\it Appendix C: Model of interference curve}---We describe a minimal model that describes the interference curve in the presence of some experimental imperfections. The population remaining in $N=1$ and $F=1$ after optical pumping is $P_{\rm bg}$. Raman transfer couples $\ket{0}\leftrightarrow\ket{y}$ (only) with efficiency $\chi_1$ in the splitter and $\chi_2$ in the recombiner. Coherences between $F=0$ and $F=1$ are neglected. The populations entering the detectors are
\begin{align}
    P_0 &= (1-\chi_1)(1-\chi_2) + \chi_1\chi_2 \cos^2 \phi, \\
    P_1 &= \chi_1 + \chi_2 -\chi_1 \chi_2 -\chi_1 \chi_2 \cos^2\phi.
\end{align}
The detectors measure $P_1$ in ${\cal D}_A$ and $P_0$ in ${\cal D}_B$. Microwave leakage results in some $P_0$ accidentally detected in ${\cal D}_A$, and incomplete optical cycling results in some $P_1$ accidentally detected in ${\cal D}_B$. These cross-couplings are denoted by ${\rm x}_{A,B}$. Thus, the signals in the two detectors are
\begin{align}
    S_A &= (P_0 {\rm x}_A + P_1 + P_{\rm bg})\epsilon_A, \\
    S_B &= (P_0 (1-{\rm x}_A) + P_1 {\rm x}_B)\epsilon_B.
\end{align}

With these definitions, and eliminating terms that are second-order in the small quantities ${\rm x}_A$, ${\rm x}_B$, $P_{\rm bg}$, the asymmetry ${\cal A}=(S_B - \frac{\epsilon_B}{\epsilon_A} S_A)/(S_B+ \frac{\epsilon_B}{\epsilon_A}S_A)$ can be written
\begin{equation}
    {\cal A}(\phi)={\cal B} + {\cal C}\cos(2\phi) + {\cal D}\cos(4\phi),
    \label{eq:interference_model}
\end{equation}
where 
\begin{equation}
    {\cal B}=\frac{1\!-\!2{\rm x}_A\!-\!(2\bar{\chi}\!-\!\frac{3}{2}\chi_1\chi_2)(2\!-\!2{\rm x}_A\!-\!{\rm x}_B)\!-\!P_{\rm bg}}{Q}+\frac{\chi_1^2\chi_2^2{\rm x}_B}{4Q^2}
     \label{eq:interference_background}
\end{equation}
is the background,
\begin{equation}
    {\cal C}=\frac{\chi_1\chi_2(2-2{\rm x}_A-{\rm x}_B)}{2Q}+\frac{\chi_1\chi_2{\rm x}_B(1-2\chi_1)}{2Q^2}
     \label{eq:interference_contrast}
\end{equation}
is the contrast of the interference term, 
\begin{equation}
    {\cal D} =\frac{\chi_1^2\chi_2^2{\rm x}_B}{4Q^2}
     \label{eq:interference_doubled}
\end{equation}
is the amplitude of a frequency-doubled term, $\phi = \mu_{\rm B} (B-B_{\rm bg}) \tau/\hbar$, $B$ is the applied magnetic field and $B_{\rm bg}$ is the background magnetic field. Here, we have introduced
$\bar{\chi}=\frac{\chi_1+\chi_2}{2}$,
and $ Q=1+{\rm x}_B(2\bar{\chi}-\frac{3}{2}\chi_1\chi_2)+P_{\rm bg}$.

Figure \ref{fig:pmt_fits}(a) shows the efficiency of the splitter, $\chi_1$, determined from PMT data equivalent to Fig.~\ref{fig:stirap}. The model shown in the figure is $\chi=\sin^2(\frac{1}{2}\int\Omega_{\rm eff}(t')dt')$ where $\Omega_{\rm eff}(t)=\Omega_0(t)\Omega_1(t)/2\Delta$ is determined from the velocity and the measured laser beam profiles. The data agrees very well with this model. Figure \ref{fig:pmt_fits}(b,c) show the contrast and offset determined from interference curves measured using the PMTs. The results are similar to those obtained from the EMCCDs shown in Fig.~\ref{fig:interference}, but have higher velocity resolution. The interference curves have high contrast and low background for molecules near the mean speed. The figure compares the fit results to the values expected from Eqs.~(\ref{eq:interference_contrast}) and (\ref{eq:interference_background}) along with independent measurements of $\chi_{1,2}$, ${\rm x}_{A,B}$, and $P_{\rm bg}$. We find very good agreement for $v \ge 140$~m/s, but some disagreement at lower speeds where the interference curves have higher quality than anticipated. This may be due to drifts between measurements of the interference curves and of the auxiliary parameters, which have a larger effect at small $v$ where the number of molecules is small.

The main paper presents interference curves taken with the EMCCDs. Here, the time bins are wide enough that we need to account for the distribution of $\tau$ values, $p(\tau)$, contributing to each time bin. The resulting model is
\begin{equation}
    f(B)=\int_{\tau_0-\delta\tau/2}^{\tau_0+\delta\tau/2} {\cal A}(\phi)\,p(\tau)\,d\tau,
    \label{eq:interference_model_averaged}
\end{equation}
where $\tau_0$ and $\delta \tau$ are the centre and width of $\tau$ values within the time bin. Since $\tau$ is proportional to the arrival time at the detectors, $p(\tau)$ is determined from the time-resolved PMT data. We have taken $p(\tau)$ to be uniform within each time bin, and fixed $\delta \tau$ to the value determined from the bin width. We also fixed ${\cal D}=0$ since allowing it to float did not improve the fits. In this case, Eq.~(\ref{eq:interference_model_averaged}) evaluates to
\begin{equation}
    f(B)={\cal B}+{\cal C}\frac{1}{\delta\phi}\cos(2\phi_0)\sin(\delta\phi),
    \label{eq:interference_model_averaged_evaluated}
\end{equation}
where $\phi_0 =\mu_{\rm B} (B-B_{\rm bg}) \tau_0/\hbar$ and $\delta\phi = \mu_{\rm B} (B-B_{\rm bg}) \delta\tau/\hbar$.

\end{document}